\documentclass[manuscript]{acmart}

\AtBeginDocument{%
  \providecommand\BibTeX{{%
    \normalfont B\kern-0.5em{\scshape i\kern-0.25em b}\kern-0.8em\TeX}}}

\setcopyright{acmcopyright}
\copyrightyear{2022}
\acmYear{2022}
\acmDOI{10.1145/3490099.3511155}

\acmConference[IUI '22]{27th International Conference on Intelligent User Interfaces}{March 22--25, 2022}{Helsinki, Finland}
\acmPrice{15.00}
\acmBooktitle{27th International Conference on Intelligent User Interfaces (IUI '22), March 22--25, 2022, Helsinki, Finland}
\acmISBN{978-1-4503-9144-3/22/03}

\usepackage{amsmath,amsfonts,bm}

\def\eqref#1{equation~\ref{#1}}
\def\1{\bm{1}}

\DeclareMathAlphabet{\mathsfit}{\encodingdefault}{\sfdefault}{m}{sl}
\SetMathAlphabet{\mathsfit}{bold}{\encodingdefault}{\sfdefault}{bx}{n}

\newcommand{\KL}{D_{\mathrm{KL}}}

\newcommand{\shortn}{\textup{\texttt{-}}}

\newcommand{\name}{Tumera+}

\usepackage{bm}
\usepackage{wrapfig}
\usepackage{makecell}

\begin{document}

\title{Interpretable Aesthetic Analysis Model for Intelligent Photography Guidance Systems}


\author{Xiaoran Wu}
\affiliation{%
  \institution{The Dept of Computer Science and Technology, Tsinghua University}
  \city{Beijing}
  \country{China}
}

\renewcommand{\shortauthors}{Xiaoran Wu}

\begin{abstract}
An aesthetics evaluation model is at the heart of predicting users' aesthetic experience and developing user interfaces with higher quality. However, previous methods on aesthetic evaluation largely ignore the interpretability of the model and are consequently not suitable for many human-computer interaction tasks. We solve this problem by using a hyper-network to learn the overall aesthetic rating as a combination of individual aesthetic attribute scores. We further introduce a specially designed attentional mechanism in attribute score estimators to enable the users to know exactly which parts/elements of visual inputs lead to the estimated score. We demonstrate our idea by designing an intelligent photography guidance system. Computational results and user studies demonstrate the interpretability and effectiveness of our method.
\end{abstract}




\maketitle

\section{Introduction}

Aesthetic experience has been celebrated for evoking unusual cognitive and affective emotions for thousands of years~\cite{kant2000critique,dewey1958experience,goodman1976languages}, and it is not surprising that aesthetics is closely related to our appreciation of the interaction with computers~\cite{tractinsky1997aesthetics,tullis1981evaluation,aspillaga1991screen,toh1998cognitive,szabo1999effects,heines1984screen,grabinger1993computer}. As a result, predicting users' aesthetic experience becomes a critical component for many real-world human-computer interaction applications, such as user interface design~\cite{ngo2003modelling}, image retrieval~\cite{westerman2007creative}, photography guidance~\cite{wu2021tumera}, image cropping~\cite{wang2017deep}, photo enhancement~\cite{bhattacharya2010framework}, photo management~\cite{datta2007learning}, etc. Central to such a prediction model is an aesthetic quality evaluation algorithm estimating the aesthetic quality for the visual input.

Some previous works base their aesthetic evaluation models on hand-crafted features. For example, image quality is estimated according to basic visual features like global edge distribution~\cite{ke2006design}, color distribution~\cite{ke2006design,aydin2014automated}, blurriness~\cite{tong2004classification}, sharpness~\cite{aydin2014automated}, depth~\cite{aydin2014automated}, and saliency~\cite{wu2010good,tang2013content,zhang2014fusion}, while features like overall density, local density, grouping, and layout complexity~\cite{tullis1997screen,streveler1984quantitative,ngo2003modelling} are utilized to evaluate user interfaces aesthetically. Hand-crafted features exploit domain and expert knowledge, but it is largely unclear whether these features are effective and powerful enough for modelling the complicated and intertwined human aesthetic standards.

To automatically extract features with a high representational capacity for aesthetic evaluation, deep learning~\cite{lecun2015deep} methods have been extensively studied. These methods typically include a convolutional neural network as a learnable feature extractor, followed by several fully-connected layers for aesthetic score prediction. Researchers design special schemes such as multi-column architecture~\cite{lu2015rating,zhang2016describing,kao2016hierarchical}, multi-task training~\cite{kong2016photo,wang2016brain}, and adaptive spatial pooling~\cite{mai2016composition} and obtain state-of-the-art prediction performance on large-scale aesthetics testbeds~\cite{murray2012ava,kong2016photo}.

Despite these achievements, existing methods largely lack interpretability and are more like a black box directly mapping inputs to aesthetic scores. These methods may be sufficient for applications like recommendation but become especially problematic when it comes to some human-computer interaction applications, where users care more about \emph{why} they get the score and how they can improve their work accordingly. For example, a previous photography guidance system~\cite[Tumera]{wu2021tumera} evaluates the aesthetic quality of current images in the viewfinder, giving scores for different aesthetic attributes like composition and color vividness. A low score reminds users of poor photo composition or color but contains little information about how to improve it. Similarly, a novice user interface designer may wonder how to redesign the interface when getting feedback saying that the density of the layout can be improved.

Closing this gap between current methods and the expectation is challenging. Detailed suggestions require fine-grained aesthetic analysis of the visual input. We pinpoint the following drawbacks of existing methods that prevent interpretability. First, current methods predict overall and attribute (lighting, composition, etc.) aesthetic scores separately~\cite{kong2016photo,ngo2003modelling} and cannot judge which attributes contribute more to the overall aesthetic evaluation. Users may focus on minor shortcomings when refining their work. Second, to improve a specific aesthetic attribute, users need to know how different elements or regions in the input contribute to the attribute score. Existing methods cannot provide such information. Consequently, users may modify the wrong part without improving the aesthetic quality.

In this paper, we solve these challenges by proposing an interpretable aesthetic evaluation model. Novelties include: (1) We propose a learnable linear combination and decompose the overall aesthetic score as a combination of individual attribute scores. The advantage is that each attribute will be given a weight quantifying how much it contributes to overall aesthetic quality. Given that different attributes are typically intertwined, we generate the parameters of the linear decomposition by a hyper-network~\cite{ha2016hypernetworks} conditioned on the global image, which automatically learns the interrelationship between attributes by back-propagation. (2) An attentional module is introduced into the attribute score predictor. In this way, predictors attend to different input regions when evaluating different attributes. The attention mask highlights regions or elements that significantly influence the evaluation result. We further optimize the mutual information between attended regions and attribute scores to obtain more a more accurate attentional model.

We demonstrate our idea by designing an interactive photography guidance system, \name. \name~users get timely feedback during shooting reminding them of the most significant shortcomings of current photos. Aesthetic scores, detailed explanations, and suggestions on how to improve will be given when requested. We provide visualization results to demonstrate that our method enables aesthetic interpretability by synergizing feature extraction, feature attention, and aesthetics regression in a unified model. User studies are carried out and proves that the enhanced interpretability largely improves the quality of photography taken by users.
\section{Related Works}

Automatically assessing image aesthetic promote many real-world applications like image retrieve~\cite{westerman2007creative}, photography guidance~\cite{wu2021tumera}, image cropping~\cite{wang2017deep}, photo enhancement~\cite{bhattacharya2010framework}, photo management~\cite{datta2007learning}, etc. Computationally modelling the process of human aesthetic experience is non-trivial. Challenges in judging the quality of an image include~\cite{deng2017image}: (1) computationally analysing and modelling the entangled photographic rules, (2) knowing the aesthetic differences in different images types, such as portrait or scenery, and in different shooting techniques, such as depth of field and high-dynamic range, (3) obtaining a large amount of human-annotated data for training or testing.

Conventional approaches try to solve these problems with \textbf{handcrafted features}. Unsurprisingly, the design of these features requires a large amount of domain knowledge and engineering skill. Hand-designed features generally belong to one of the following four categories. (1) Basic visual features, including global edge distribution~\cite{ke2006design}, color distribution~\cite{ke2006design,aydin2014automated}, hue count~\cite{ke2006design}, low-level contrast~\cite{tong2004classification,ke2006design} and brightness~\cite{ke2006design}, blurriness~\cite{tong2004classification}, sharpness~\cite{aydin2014automated}, depth~\cite{aydin2014automated}. Related to our work, previous research introduces visual attention mechanisms into image aesthetics evaluation~\cite{sun2009photo,you2009perceptual}. However, they apply attention to image saliency and do not model or provide interpretability to other aesthetic attributes and the overall aesthetic score. Aforementioned visual features can also be extracted locally from image regions that are potentially the most attractive to humans~\cite{luo2008photo,wong2009saliency,nishiyama2011aesthetic,lo2012statistic}. (2) Composition features. Composition is typically related to the position of a salient object in an image. To make an object salient, one can adopt techniques like the rule of third, low depth of field, and opposing colors. Various handcrafted features~\cite{bhattacharya2010framework,dhar2011high,bhattacharya2011holistic,wu2010good,wu2010good,tang2013content,zhang2014fusion} are designed to model this aesthetic aspect. (3) General-purpose features. These features are not specifically designed for aesthetic evaluation, such as scale-invariant feature transform (SIFT)~\cite{yeh2012relative}, bag of word (BOV) words, and Fisher vector (FV)~\cite{marchesotti2011assessing,marchesotti2013learning,murray2012ava}. (4) Task-specific features. These features are designed specifically for a category of images. For example,~\citet{li2010aesthetic} and~\citet{lienhard2015low} design features for photos with faces, \citet{su2011scenic} and~\citet{yin2012assessing} study landscape photos, and \citet{sun2015aesthetic} evaluate the visual quality of Chinese calligraphy. 

Based on these features, heuristic rules~\cite{aydin2014automated}, SVMs~\cite{datta2006studying,luo2008photo,wong2009saliency,wu2010good,nishiyama2011aesthetic,dhar2011high,wu2010good,yin2012assessing,tang2013content,lienhard2015low}, SVR~\cite{bhattacharya2010framework,li2010aesthetic,bhattacharya2011holistic}, naive Bayes classifiers~\cite{ke2006design}, boosting~\cite{tong2004classification,su2011scenic}, regression~\cite{sun2009photo}, and Gaussain mixture model~\cite{marchesotti2011assessing,marchesotti2013learning,murray2012ava} are used to get the overall aesthetic score estimation. For detailed discussion of these methods, we refer readers to~\citet{deng2017image}.

\textbf{Deep aesthetic models}. \citet{wang2016multi} modify the AlexNet architecture for aesthetic evaluation. Specifically, the authors replace the fifth convolutional layer of AlexNet by a group of seven convolutional layers. The purpose is to better extract scene-specific features. Each of the seven convolutional layers is dedicated to one category of scene. The output from these layers are averaged and the whole architecture is trained end-to-end with backpropagation. 

\citet{tian2015query} train an aesthetic model with a smaller fully-connected layer. They expect the deep model to extract useful features which can be classified by an SVM. Typical deep learning models require the input to have the same size. Therefore, images are usually resized or cropped before fed into the network. Visual features related to image aesthetics may be corrupted during this process. To solve this problem,~\citet{lu2015deep} aggregate features of randomly selected image patches that are of the same size. Patches are directly fed into the CNN, so that the original aesthetic information is retained. Another attempt to deal with this information loss problem is adaptive spatial pooling~\cite{mai2016composition}. Another contribution of~\cite{mai2016composition} is to adopt a multi-column CNN structure to deal with images with different scenes. Multi-column neural networks have also been used in aesthetics evaluation for global-local feature combination~\cite{lu2015rating}, patch-global input tradeoff~\cite{zhang2016describing}, and separate object-scene-texture processing~\cite{kao2016hierarchical}.

\citet{kong2016photo} present a dataset containing around 10,000 images, each with an overall aesthetic score and scores for 11 aesthetic attributes. The authors also propose a Siamese architecture which takes two images as input and outputs a rank of the images. It is shown that such a training scheme can help distinguish images with similar aesthetic quality. Our work is orthogonal to the aforementioned works involving deep learning. One can easily integrate our methods into these model and expect enhanced interpretability. On the other hand, introducing these models into our framework can further improve the accuracy of the proposed learning architecture.

Similar to our work,~\citet{wang2016brain} predict different style attributes and use them as the input to a final CNN for predicting the overall score distribution. The difference between this work and ours lies in three aspects. (1) We use a hyper-network to incorporate global information of the image when predicting the overall score, which makes a better use of the interrelationship among attributes. (2) We directly predict attribute scores instead of features to better use label information. (3) We use an attentional mechanism when predicting the attribute scores, improving the interpretability of our model for better interaction experience.

Apart from image aesthetic evaluation, it has been found that interface aesthetics greatly influence system acceptability~\cite{tractinsky1997aesthetics,kurosu1995apparent}, comprehensibility~\cite{tullis1981evaluation,tullis1984predicting}, and productivity~\cite{kelster1983making,aspillaga1991screen,toh1998cognitive,szabo1999effects,heines1984screen,grabinger1993computer}. An aesthetic metric would be an essential aid of screen design. Previous work~\cite{tullis1997screen,streveler1984quantitative,ngo2003modelling} propose hand-crafted metrics, such as overall density, local density, grouping, and layout complexity, for alphanumeric displays. \citet{sears1993layout} proposes the concept of layout User appropriateness to assess whether the spatial layout is in harmony with the task.~\citet{rahman2021ruite,gupta2020layout} use a transformer for refining user interface layout aesthetics.~\citet{dou2019webthetics} use CNNs for predicting website aesthetics. 
\section{Aesthetic evaluation model}

The major aim of our work is to enhance the interpretability of aesthetic evaluation models for a better user interaction experience. To this end, we focus on two novel functions of the evaluation model. (1) We propose to learn the overall aesthetic score as an input-dependent decomposition of individual attribute scores so that users know which attributes contribute the most to the overall aesthetic quality of the image. (2) For each attribute, we learn an attentional mask and let the model attend different input regions.Based on the learned attention mechanism, our model provides detailed explanations and suggestions accompanying the aesthetic evaluation.

\begin{figure}
    \centering
    \includegraphics[width=\linewidth]{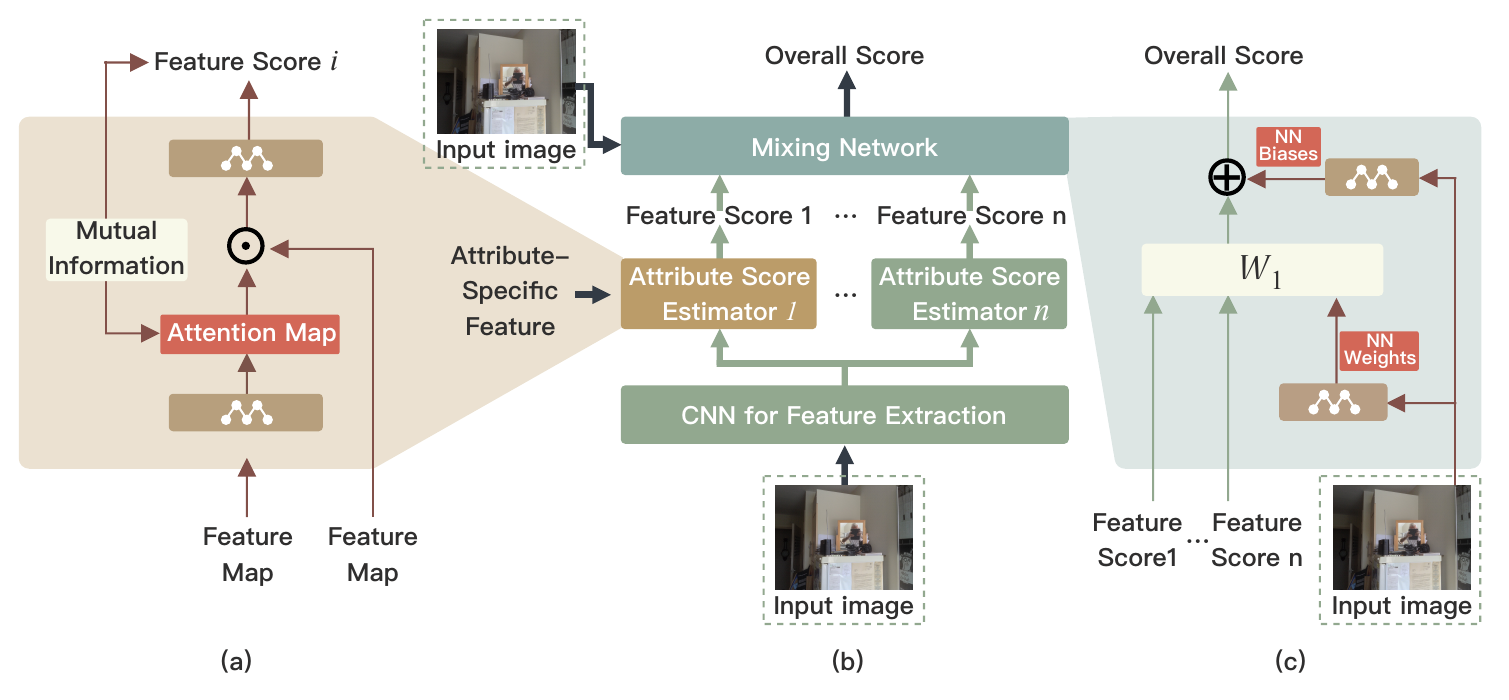}
    \caption{Network architecture of our aesthetics evaluation framework. (a) Attribute score predictor. (b) Overall framework. (c) Aesthetic decomposition network.}
    \label{fig:framework}
\end{figure}

In this paper, we focus on the aesthetic assessment of images, but our model can be easily extended to other applications, like user interface design. Fig.~\ref{fig:framework} shows the architecture of our learning framework.

The proposed framework (Fig.~\ref{fig:framework} (b)) consists of three modules. The first part is a convolutional neural network (CNN) for feature extraction. Since we usually have a limited number of training samples with human-labeled aesthetic scores, it is not practical to learn the feature extractor from scratch. Thus we start from a pre-trained ResNet~\cite{he2016deep} for its proven ability to learn robust image representations that show ever-improving performance in the task of aesthetic evaluation~\cite{mccormack2020understanding,jang2021analysis,mccormack2021deep}. Feature maps extracted by the CNN are fed into attribute score estimators ,which output attribute scores. At the heart of these attribute estimators is a specially designed attentional mechanism that provides better interpretability. Estimated attribute scores are combined with a decomposition network to generate the overall aesthetic evaluation. The whole structure is differentiable and can be trained in an end-to-end fashion. In the following sections, we introduce the detailed structures for attribute score estimators and the decomposition network and also discuss the training scheme of our model. 

\subsection{Attribute Score Estimator}
The aim of an attribute score estimator is to predict a score for one of the aesthetic attributes. In this paper, we consider eleven aesthetic attributes: $\mathtt{elements\ balance}$, $\mathtt{color\ harmony}$, $\mathtt{content}$, $\mathtt{depth\ of\ field}$, $\mathtt{light}$, $\mathtt{motion\ blur}$, $\mathtt{object}$, $\mathtt{repetition}$, $\mathtt{rule\ of\ thirds}$, $\mathtt{symmetry}$, and $\mathtt{color\ vividness}$. 

The major novelty here is (1) an attention module enabling the model (and the user) to attend different regions or elements when considering different attributes and (2) the way to train the attentional module. A separate attention module is trained for each of the attributes. The attention module is a fully-connected network $f_{\text{att}}(\ \cdot\ ; \theta_{\text{att}})$ with SoftMax activation at the last layer, parameterized by $\theta_{\text{att}}$, takes the feature map $z_{\text{img}}$ generated by the feature extractor as input and outputs an attention mask $z_{\text{att}}$, which has the same size as $z_{\text{img}}$. The attention mask and the feature map are combined together by an element-wise multiplication. This attentional weighted feature map is fed into another network $f_{\text{attr}}(\ \cdot\ ; \theta_{\text{attr}})$ to predict the attribute score. Prediction losses are used to train the attention modules end-to-end (Sec.\ref{sec:loss}).

Gradients induced by prediction losses may not contain much supervision information to get accurate attention maps. We thus introduce an information-theoretical regularizer to maximize the mutual information between attentional masks and the corresponding attribute scores:
\begin{equation}
    MI(A,S|M) = \int\int\int p(a,s,m) \log\frac{p(a,s|m)}{p(a|m)p(s|m)} dadsdm,
\end{equation}
where $A$, $S$, and $M$ are the random variable of attentional maps, attribute scores, and input images, respectively. We have that
\begin{equation}
    MI(A,S|M) = \int\int p(m,a)\int p(s|a,m) \log\frac{p(s|a,m)}{p(s|m)} ds dadm.
\end{equation}
While $p(s|m,a)$ has been approximated by attribute score estimators, $p(s|m)$ needs to be estimated. Therefore, we introduce a variational posterior estimator $q_\phi$ and get a lower bound of the mutual information objective:
\begin{align}
    MI(A,S|M) &= \int\int p(m,a)\int p(s|a,m) \log\frac{p(s|a,m)}{q_\phi(s|m)} ds dadm + \int p(m)\int p(s|m) \log\frac{q_\phi(s|m)}{p(s|m)} ds dm \nonumber \\
    &= \mathbb{E}_{m\sim \mathcal{T}, a\sim f_{\text{att}}(m) }\left[\KL[p(s|a,m)\|q_\phi(s|m)]\right] + \mathbb{E}_{m\sim \mathcal{T}}\left[\KL[q_\phi(s|a)\|p(s|a)]\right]  \\
    & \ge \mathbb{E}_{m\sim \mathcal{T}, a\sim f_{\text{att}}(m) }\left[\KL[p(s|a,m)\|q_\phi(s|m)]\right]\nonumber,
\end{align}
where the inequality holds because of the non-negativity of the KL divergence. The expectation operator here means that images are sampled from the training set $\mathcal{T}$ and the attentional maps are generated by the attention module $f_{\text{att}}$. Intuitively, this lower bound quantifies the difference of attribute scores with and without attentional maps, highlighting the influence of attentional masks.

To calculate the KL divergence, the output of the attention module needs to be a distribution over scores. To this end, we introduce a Gaussian distribution
\begin{equation}
p(s|a,m)=\mathcal{N}(f_{\text{attr}}(f_{\text{att}}(z_{\text{img}})), \Sigma),\label{equ:distribution}
\end{equation}
where $\Sigma$ is a constant covariance matrix. In Sec.~\ref{sec:loss}, we describe the training approach in detail.

\subsection{Aesthetic Decomposition Network}
To learn how much each attribute contributes to the overall aesthetic score, we propose to learn the overall score as a combination of individual attribute scores. Specifically, we propose to learn
\begin{equation}
    s_{\text{aes}} = \sum_i w_i * s_{\text{att}_i},
\end{equation}
where $s_{\text{aes}}$ is the overall quality score, and $s_{\text{att}_i}$ is the score of the i-$th$ attribute. In this way, the overall score is represented as a linear combination of attribute scores. The question is how can we get the weights $w_i$ of this linear combination. Intuitively, the interrelationship between attributes is determined by the structure of the original image. Therefore, we adopt a hyper-network~\cite{ha2016hypernetworks} to learn the linear combination parameters. This network is called "hyper" because it generates the parameters of another function. Specifically, a network $f_{\text{mix}\shortn w}(\ \cdot\ ; \theta_{mix\shortn w})$ with a SoftMax activation after the last layer processes the input image and outputs a vector $\bm w=\langle w_1, w_2, \dots, w_K\rangle$ whose elements serve as the weights of the linear combination.

It is worth noting that gradients can flow through $\bm w$ into the hyper-network, so that its parameters can be updated. On the other hand, gradients can also flow into attribute score estimators through $s_{\text{att}_i}$. Therefore, the use of hyper-networks does not influence the end-to-end differentiability of the whole learning framework. 

\subsection{Training Scheme}\label{sec:loss}
Our model is trained by minimizing three losses. The first loss is to minimize the mean square error between the predicted aesthetic quality $s_{\text{aes}}$ and the groundtruth $y_{\text{aes}}$:
\begin{equation}
    \mathcal{L}_{\text{aes}}(\theta_{\text{att}}, \theta_{\text{attr}}, \theta_{\text{mix}\shortn w}) = \mathbb{E}_{m\sim\mathcal{T}}\left[y_{\text{aes}}(m) - s_{\text{aes}}(m;\theta_{\text{att}}, \theta_{\text{attr}}, \theta_{\text{mix}\shortn w}) \right]^2,\label{equ:loss_aes}
\end{equation}
where the expectation operator means that images $m$ are sampled from the training set $\mathcal{T}$. Similarly, the loss for minimizing the attribute score prediction error is given by:
\begin{equation}
    \mathcal{L}_{\text{att}}(\theta_{\text{attr}}, \theta_{\text{att}}) = \mathbb{E}_{m\sim\mathcal{T}}\left[\sum_i y_{\text{aes}_i}(m) - s_{\text{aes}_i}(m;\theta_{\text{att}}, \theta_{\text{attr}}) \right]^2,\label{equ:loss_att}
\end{equation}
where $y_{\text{aes}_i}$ is the groundtruth score of the i-$th$ attribute provided by the dataset. In Eq.~\ref{equ:loss_aes} and~\ref{equ:loss_att}, attribute scores are sampled from the distribution $p(s|m,a)$ (Eq.~\ref{equ:distribution}). We use the reparameterization trick~\cite{kingma2014stochastic} to sample scores so that our framework is end-to-end differentiable.

For maximizing the mutual information objective, we minimize the following loss:
\begin{equation}
        \mathcal{L}_{MI}(\theta_{\text{attr}},\theta_{\text{att}},\phi) = -\mathbb{E}_{m\sim \mathcal{T}, a\sim f_{\text{att}}(m) }\left[\KL[p(s|a,m)\|q_\phi(s|m)]\right].
\end{equation}

By introducing scaling factor $\lambda_{\text{att}}$ and $\lambda_{MI}$, the total loss for training our model is:
\begin{equation}
    \mathcal{L}(\theta_{\text{att}}, \theta_{\text{attr}}, \theta_{\text{mix}\shortn w},\phi) = \mathcal{L}_{\text{aes}}(\theta_{\text{att}}, \theta_{\text{attr}}, \theta_{\text{mix}\shortn w}) + \lambda_{\text{att}}\mathcal{L}_{\text{att}}(\theta_{\text{attr}}, \theta_{\text{att}}) + \lambda_{MI}\mathcal{L}_{MI}(\theta_{\text{attr}},\theta_{\text{att}},\phi). \label{equ:total_loss}
\end{equation}
% \begin{figure}
%     \centering
%     \includegraphics[width=0.9\linewidth]{fig/interface2.pdf}
%     \caption{Caption}
%     \label{fig:interface-1}
% \end{figure}
\section{Demo Application: Intelligent Photography Assistant}

\begin{wrapfigure}[21]{r}{0.8\linewidth}
    \vspace{-2em}
    \includegraphics[width=\linewidth]{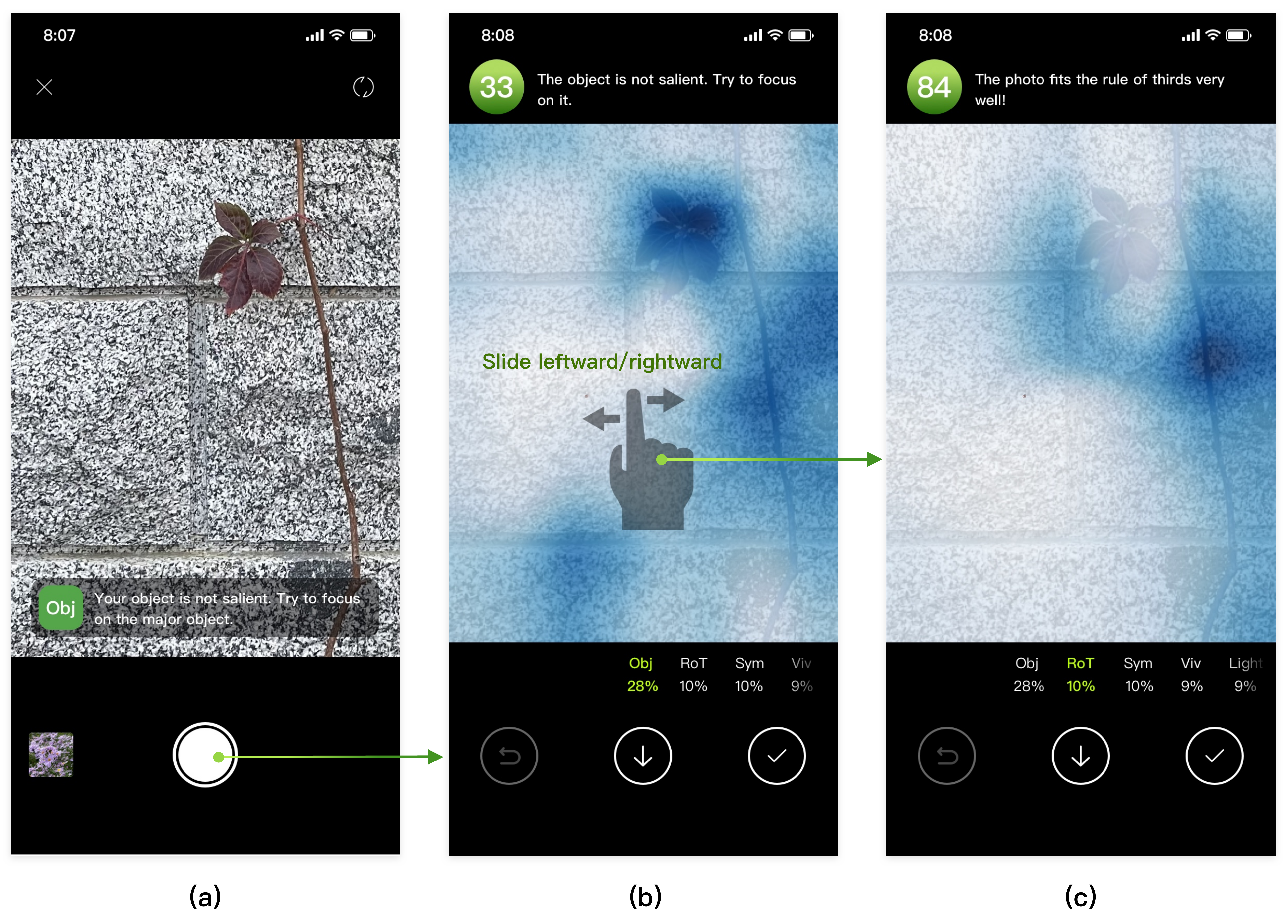}
    \caption{User interfaces with (a) real-time prompts and (b-c) detailed evaluations, explanations, and suggestions for photographs.}
    \label{fig:interface-1}
\end{wrapfigure}
We apply our aesthetic model to an interactive photography guidance system. We also notice that our model can be used for other tasks like user interface design but leave it for future work. In this section, we describe our intelligent user interface for photography guidance. Based on the proposed aesthetic evaluation model, we design heuristics and provide users with brief prompts or detailed scores and suggestions according to their needs. Specifically, we design the following interfaces to provide users with various photography guidance.

\subsection{Real-Time Prompts}
Real-time prompts are immediate feedback provided during shooting (Fig.~\ref{fig:interface-1} left). The proposed aesthetic evaluation model runs asynchronously with the camera, requesting the image in the viewfinder every 0.5 second and evaluating their aesthetic quality. Based on the evaluation, our system prompts users to improve their photography by pointing out the disadvantages of current images. Trying not to be distracting, we select the attribute with the largest weight among all attributes whose score is below the overall score and show the information in the form of pop-up messages. 

Messages are based on attribute scores and the image region indicated by the attention map. For example, for the attribute $\mathtt{object}$, if its score is below the overall score, we calculate the area to which the attentional module attends. If the area exceeds one-half of the entire image, the prompt would be \emph{Your object is too salient. Try to make it small}. Otherwise, if the area is less than one-ninth of the entire image, the prompt would be \emph{Your object is not salient. Try to focus on the major object}. For attribute $\mathtt{color\ vividness}$, we translate the RGB values within the attentional map into HSV values and calculate the average chrominance. If the chrominance is too low, the prompt would be \emph{Highlighted color is not vivid}. If the chrominance is too high, the prompt would be \emph{Highlighted color is too vivid}. Similarly, we design heuristics for all the other attributes.

We note than such guidance is only possible with the learned attentional module. Otherwise, we only know whether an attribute contribute positively or negatively to the overall quality, but cannot give accurate messages with detailed explanations. 

\subsection{Detailed Feedback and Suggestions}
Real-time prompts only give hints regarding the attribute that needs the most urgent attention, and users are unaware of the overall quality of the image as well as the information about other attributes. To makeup for this shortage, we provide detailed feedback and suggestion by designing the following interaction interfaces.

(1) Detailed suggestions. When the user presses the shutter button, the current image is sent to the evaluation model. Based on the returned results, we provide detailed feedback and suggestions. Specifically, we sort the attributes according to their weights. Then, users can slide their window leftward or rightward (Fig.~\ref{fig:interface-1} (b-c)) to see scores, attentional maps, and corresponding suggestions associated with different attributes. The same as real-time prompts, detailed suggestions are given by heuristics based on attribute scores and attentional maps. Users can better understanding the suggestions by looking at the attentional maps. 

\begin{wrapfigure}[18]{r}{0.5\linewidth}
    \vspace{-2em}
    \includegraphics[width=\linewidth]{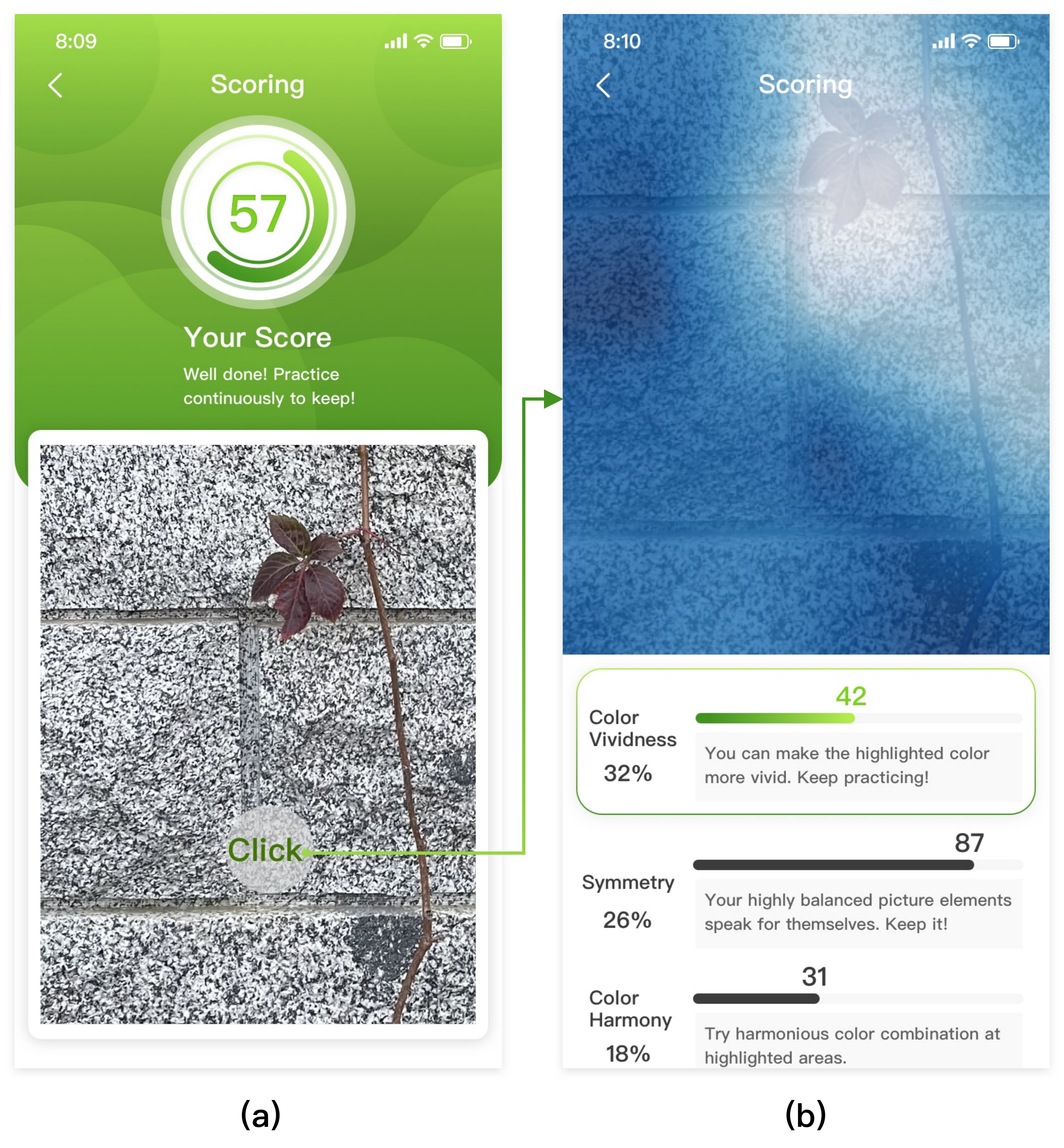}
    \caption{User interfaces with regional aesthetic evaluation and improvement suggestions.}
    \label{fig:interface-2}
\end{wrapfigure}
(2) Regional suggestions. To adjust their camera and get a better photo, users may need some regional evaluation. For example, in Fig.~\ref{fig:interface-2}, a user may want to know whether she should include the wall in the picture. To this end, we design a clicking interaction function. After a user clicks at a specific region of the image, we calculate the average attentional value within the region for each attribute. These average values are sorted so that we know which attributes are attending to this region. If more attributes with a large average attentional value contribute negatively to the overall photo quality, we would suggest the user to remove the region when they shoot the same scene again. 
\begin{figure}
    \centering
    \includegraphics[width=\linewidth]{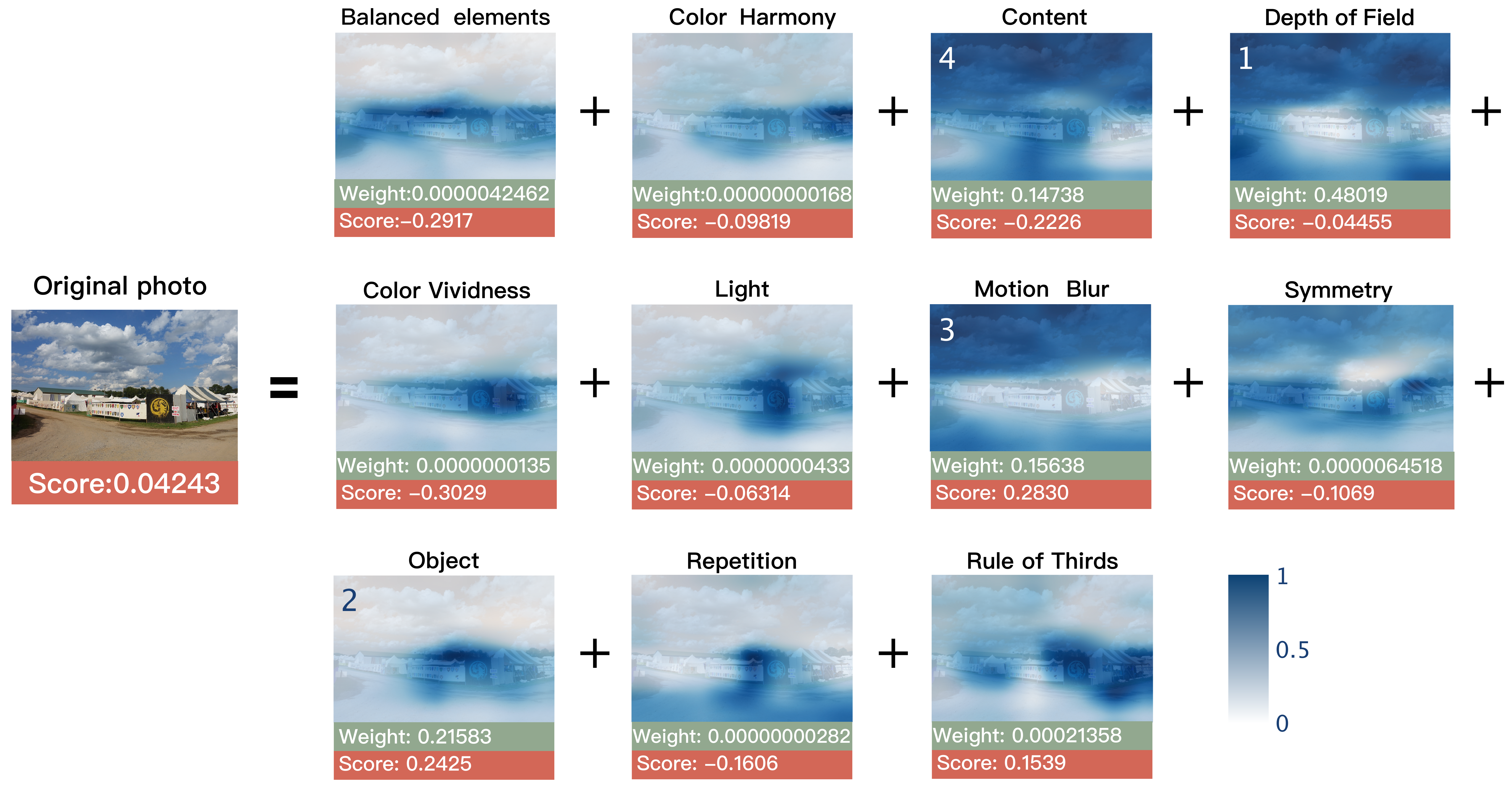}
    \caption{In our model, attribute score estimators attend to different regions of the input image, giving a score evaluating corresponding attributes. Scores are mixed linearly to generate the overall score. The parameters of the linear combination depend on the input image. Scores shown are in the range $[-1,1]$, where an $1$ stands for the highest aesthetic quality.}
    \label{fig:combination}
\end{figure}

\section{Experiments}\label{sec:exp-com}

In this section, we design experiments to answer the following two questions: (1) Whether our model can learn interpretable aesthetic evaluations. (Sec.~\ref{sec:exp-int}) (2) Does the learned evaluation model align with judgements of photography experts. (Sec.~\ref{sec:exp-com-exp}) In Sec.~\ref{sec:exp-user}, we carry out user studies to test whether our photography guidance system can help improve the quality of users' photos.  

\subsection{Image Aesthetic Datasets}
We train our model on the public image aesthetic datasets AADB~\cite{kong2016photo}. This dataset includes 10,000 images in total. Each image in the dataset has an overall aesthetic quality rating and attribute assessments provided by five different individual raters. These attribute assessments cover the 11 attributes considered in this paper. Original images are presented in this dataset with different sizes. To fit our model, we resize the images to $256\times 256\times 3$. Although we can use previous methods~\cite{lu2015deep, mai2016composition} to avoid image resizing and preserve more original aesthetic-related image features, our focus in this paper is on interpretability and we leave these for future work. 

\subsection{Training Setups}
We set the loss weight $\lambda_{\text{att}}$ to $0.09$ and $\lambda_{MI}$ to $0.001$ and train the model for $100$ epochs with a mini-batch of size $32$. Optimization is carried out by an Adam optimizer~\cite{kingma2014adam} with a learning rate of $4\times 10^{-4}$. An early stop mechanism is adopted to adopt model overfitting -- we stop training when the total loss (Eq.~\ref{equ:total_loss}) does not decrease in $15$ consecutive epochs. We clip gradients with a norm larger than 5 to avoid dramatic network changes. 

A pre-trained ResNet with 101 layers is used for feature extraction. We do not use the fully-connected layers of the pre-trained model because the downstream task is different. The extracted feature maps are of the size $14\times 14\times 2048$. For the attentional module, we generate the attentional map via a fully-connected neural network with a hidden layer of $2048$ units and ReLU activation. The extracted image features are also fed into the hyper-network for generating the weights of the linear transformation for individual attribute scores. The hyper-network is a three-layer fully-connected network with ReLU activation. The two hidden layers have $102$ and $19$ units, respectively.

\subsection{Interpretable Aesthetic Model}\label{sec:exp-int}
We show our model's evaluation of an image from the AADB dataset in Fig.~\ref{fig:combination}. The left column shows the original image and the overall aesthetic score (both overall and attribute scores are in the range $[-1,1]$ in all of our experiments, but they are transformed linearly to the range $[1,100]$ in our photography guidance system) estimated by our model. The rest columns show the attention masks and scores corresponding to each aesthetic attribute. 

We can observe that the rating of this image is not very high. From the attribute weights shown in the right, we can see that four attributes---$\mathtt{depth\ of\ field}$, $\mathtt{object}$, $\mathtt{motion\ blur}$, and $\mathtt{content}$---have the most significant influence on the final score by accounting for over $99\%$ of the total weight. Among them, $\mathtt{depth\ of\ field}$ has the largest weight but contribute negatively to the overall score. The corresponding attention mask indicates which part of the image leads to the rating. As the mask highlights the  ground and sky, we know that a more salient object will improve the rating of $\mathtt{depth\ of\ field}$. The attribute $\mathtt{content}$ has the lowest score among attributes with large weights. The model mainly attends to the sky when estimating this score. The negative score and the attentional mask together suggest that the sky is not interesting and could be removed for better $\mathtt{content}$ score. 

Among the attributes with large weights, $\mathtt{object}$ and $\mathtt{motion\ blur}$ contribute positively to the final score. The attentional masks explain why the attribute scores are positive. We can see that, in the original image, the cloud is clear and free of blur, and the object texture is of good quality, which explain why the model assigns positive scores.

For other attributes with negligible weights, we can see that our model still learns meaningful attentional masks, though the corresponding attribute scores do not influence the final evaluation significantly. For example, our model attends to the salient object in this image when evaluating whether the image conforms to the rule of thirds (attribute $\mathtt{RoT}$). The attribute attentional mask sits at the horizontal line dividing the lower third of the photo from the upper two thirds.

Similar results generally hold for most pictures in the dataset. In Sec.~\ref{sec:exp-user}, we further show that our model can also generalize to photos out of the training set by showing cases encountered by our users.

\subsection{Expert Evaluation}~\label{sec:exp-com-exp}
To further test whether our model learns reasonable aesthetic evaluation metrics, we invite three photography experts, an advertising photographer, a photography graduate student, and a college photography teacher, to evaluate our model. Specifically, we show the experts with 10 pairs of photos and ask them to: (1) For each pair of image, choose the one with better overall aesthetic quality; (2) For each photo, select $5$ attributes which they think contribute the most to the overall quality; (3) For each pair of the image and each attribute, select the image with better attribute quality.

\begin{wraptable}[8]{r}{0.6\linewidth}
\vspace{-1em}
    \caption{Agreement degrees between expert evaluations and the assessments provided by our model.}
    \vspace{-1em}
    \label{tab:com-exp}
    \centering
    \begin{tabular}{cccccccc}
        \toprule
        \multicolumn{2}{c}{} &
        \multicolumn{2}{l}{Overall Score} &
        \multicolumn{2}{l}{\makecell{Attribute Scores}} &
        \multicolumn{2}{l}{\makecell{Attribute Weights}} \\
        \cmidrule(lr){1-2}
        \cmidrule(lr){3-4}
        \cmidrule(lr){5-6}
        \cmidrule(lr){7-8}
        
        \multicolumn{2}{c}{Expert \#1} & \multicolumn{2}{c}{90\%}  & \multicolumn{2}{c}{83.6\%} & \multicolumn{2}{c}{67.5\%} \\
        \cmidrule(lr){1-2}
        \cmidrule(lr){3-4}
        \cmidrule(lr){5-6}
        \cmidrule(lr){7-8}
        
        \multicolumn{2}{c}{Expert \#2} & \multicolumn{2}{c}{70\%}  & \multicolumn{2}{c}{73.6\%} & \multicolumn{2}{c}{62.5\%}  \\
        \cmidrule(lr){1-2}
        \cmidrule(lr){3-4}
        \cmidrule(lr){5-6}
        \cmidrule(lr){7-8}
        
        \multicolumn{2}{c}{Expert \#3} & \multicolumn{2}{c}{90\%}  & \multicolumn{2}{c}{85.5\%} & \multicolumn{2}{c}{82.5\%} \\
        
        \toprule
    \end{tabular}
\end{wraptable}
The expert data is then used to evaluate each of the prediction made by our model. (1) For the overall aesthetic score, we check whether the rank given by experts is the same as the one given by our model. Agreement degrees are shown in Table~\ref{tab:com-exp} (2) For the attribute weights, we check whether the most important attributes selected by our model align with the ones selected by experts. Overlap rates are shown in Table~\ref{tab:com-exp}. (3) For the attribute scores, we check whether the rank given by experts agrees with our computational model. Agreement degrees are shown in Table~\ref{tab:com-exp}. 

We can see that the agreement degree is very high in terms of the overall score, with an average of $83.3\%$ among the three experts. The prediction of attribute scores also is also well aligned with the expert evaluation, achieving an average of $80.9\%$ among experts. The prediction of the most influential attributes is more challenging than the other two prediction tasks, which is because the aesthetic attributes are typically intertwined with each other. Despite the difficulty, our model achieves an average agreement degree of $70.8\%$ with the experts. These results verify the effectiveness of our method.

\section{User Study}\label{sec:exp-user}
Having described our aesthetic model architecture and the results in terms of computational evaluation, we then carry out user studies to see whether our system can improve the aesthetic quality of photos taken by users.

We recruit $10$ volunteers to use our system. The number of male and female volunteers are equal, and they are between 21-60 years old. Since our system aims to provide guidance, volunteers are mainly beginners in photography and do not include professional photographers. Volunteers are asked to photograph a scene for two times, without and with the guidance of our system, respectively. Volunteers can freely decide which scene they want to photograph.

Table~\ref{tab:diff} shows the scores estimated by our aesthetic model for photos taken by users before and after the guidance of \name. We can see that 9 out of all 10 volunteers are able to improve photo quality with the guidance of \name. The improvements of score distribute from $5.66\%$ to $82.03\%$, with an average of $25.57\%$. These results demonstrate that our system can help users obtain photos of higher quality.

\begin{wrapfigure}[16]{r}{0.7\linewidth}
    \vspace{-1em}
    \includegraphics[width=\linewidth]{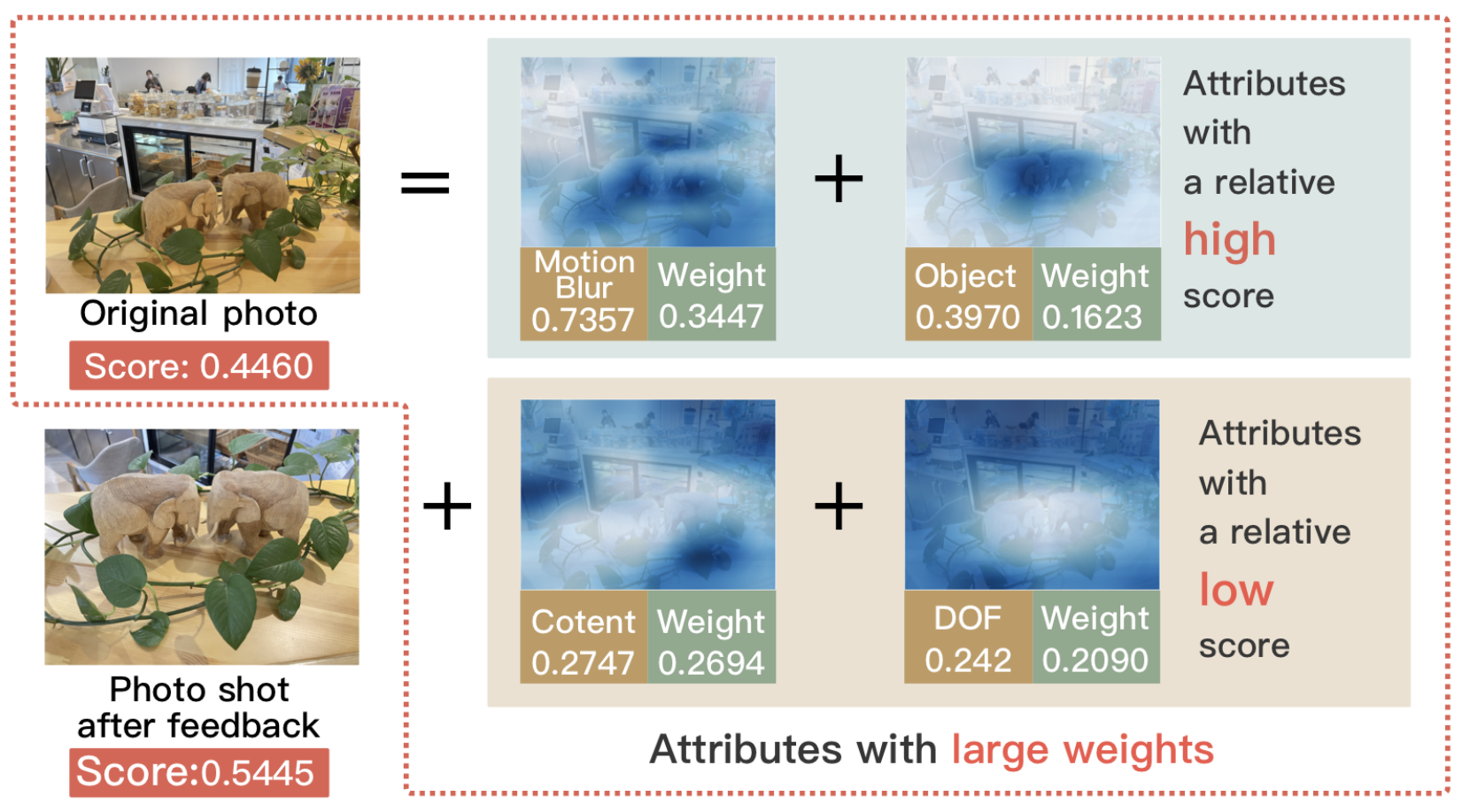}
    \caption{Comparison of photos shot before and after the guidance of our system. Suggestions provided to the user are given based on the attribute scores, the attribute weights, and the attentional mask. }
    \label{fig:diff}
\end{wrapfigure}
In Fig.~\ref{fig:diff}, we show an example pair of photos shot by one volunteer. At upper-left is the photo shot before using \name. Based on the attribute scores and attentional maps shown in the right, we give the suggestions "\emph{The highlighted content is not very interesting.}" and "\emph{Improve the depth of field by blurring or removing the highlighted area}". The suggestions, scores, and attentional maps are shown to the  volunteer. Guided by the feedback, the volunteer is able to take a photo (lower-left) of better quality.

We invite three photography experts (the same as who we invite for computational experiments in Sec.~\ref{sec:exp-com}) to evaluate whether the quality of photos indeed improves. Results are shown in the last three lines of Table~\ref{tab:diff}. We observe a great alignment between expert judgements and the evaluation given by our system. All experts agree that at least 8 photos shot after our system's guidance are better. We show the detailed results of \name~on the tested photo pairs and the corresponding  evaluations of experts and participants online\footnote{\url{https://sites.google.com/view/iui1170}}.
\begin{table}[ht]
    \centering
    \caption{Upper: Aesthetic scores before and after the guidance of our system. Lower: Whether an expert thinks that the photo taken under the guidance of \name~is better.}\label{tab:diff}
    \vspace{-1em}
    \begin{tabular}{p{1.5cm}p{0.9cm}p{0.8cm}p{0.8cm}p{0.8cm}p{0.8cm}p{0.8cm}p{0.8cm}p{0.8cm}p{0.8cm}p{0.8cm}}
    \hline
         Volunteer & 1 & 2 & 3 & 4 & 5 & 6 & 7 & 8 & 9 & 10 \\
    \hline
         Before  & \textbf{0.253} & 0.446 & 0.314 & 0.217 & 0.365 & 0.265 & 0.322 & 0.452 & 0.212 & 0.374 \\
         After   & 0.249 & \textbf{0.544} & \textbf{0.379} & \textbf{0.395} & \textbf{0.375} & \textbf{0.306} & \textbf{0.504} & \textbf{0.524} & \textbf{0.224} & \textbf{0.407} \\
         Diff.   & -1.58\% & 21.97\% & 20.70\% & 82.03\% & 2.74\% & 15.47\% & 56.52\% & 15.93\% & 5.66\% & 8.82\% \\
    \hline
         Expert \#1 & \checkmark & \checkmark & \checkmark & \checkmark & \checkmark & \checkmark & \checkmark & \checkmark & \checkmark & \checkmark \\
         Expert \#2 & \checkmark & \checkmark & \checkmark & \checkmark & & \checkmark & & \checkmark & \checkmark & \checkmark \\
         Expert \#3 & \checkmark & \checkmark & \checkmark & \checkmark & & \checkmark & \checkmark & \checkmark & \checkmark & \checkmark \\
    \hline
    \end{tabular}
\end{table}

\section{Conclusion}

In this paper, we improve visual aesthetic evaluation models by introducing interpretability so that their applicability to human-computer interaction applications is enhanced. To this end, (1) we learn to decompose the overall aesthetic score as a combination of attribute scores so that we know the contribution of each attribute; (2) we introduce a specially-designed attentional mechanism into predictors so that the contribution of different image regions and elements are explainable. We demonstrate our idea by designing an intelligent photography guidance system. Although we focus on image aesthetic evaluation in this paper, our model is expected to be extended to other tasks like user interface design. We plan to explore this direction in future work.

\bibliographystyle{ACM-Reference-Format}
\bibliography{sample-base}

\appendix

\section{Questionnaire for Expert Evaluation}
In this section, we present part of the questionnaire completed by Expert \#1. The questionnaire requests the expert to compare the two images shown in a row, select five attributes which they think are the most influential to the final aesthetic evaluation, and how these attributes contribute (positively or negatively) to the overall aesthetic quality. These results are used to calculate the agreement degree between the expert evaluations and the assessments provided by our system.

\begin{figure}
    \centering
    \includegraphics[width=\linewidth]{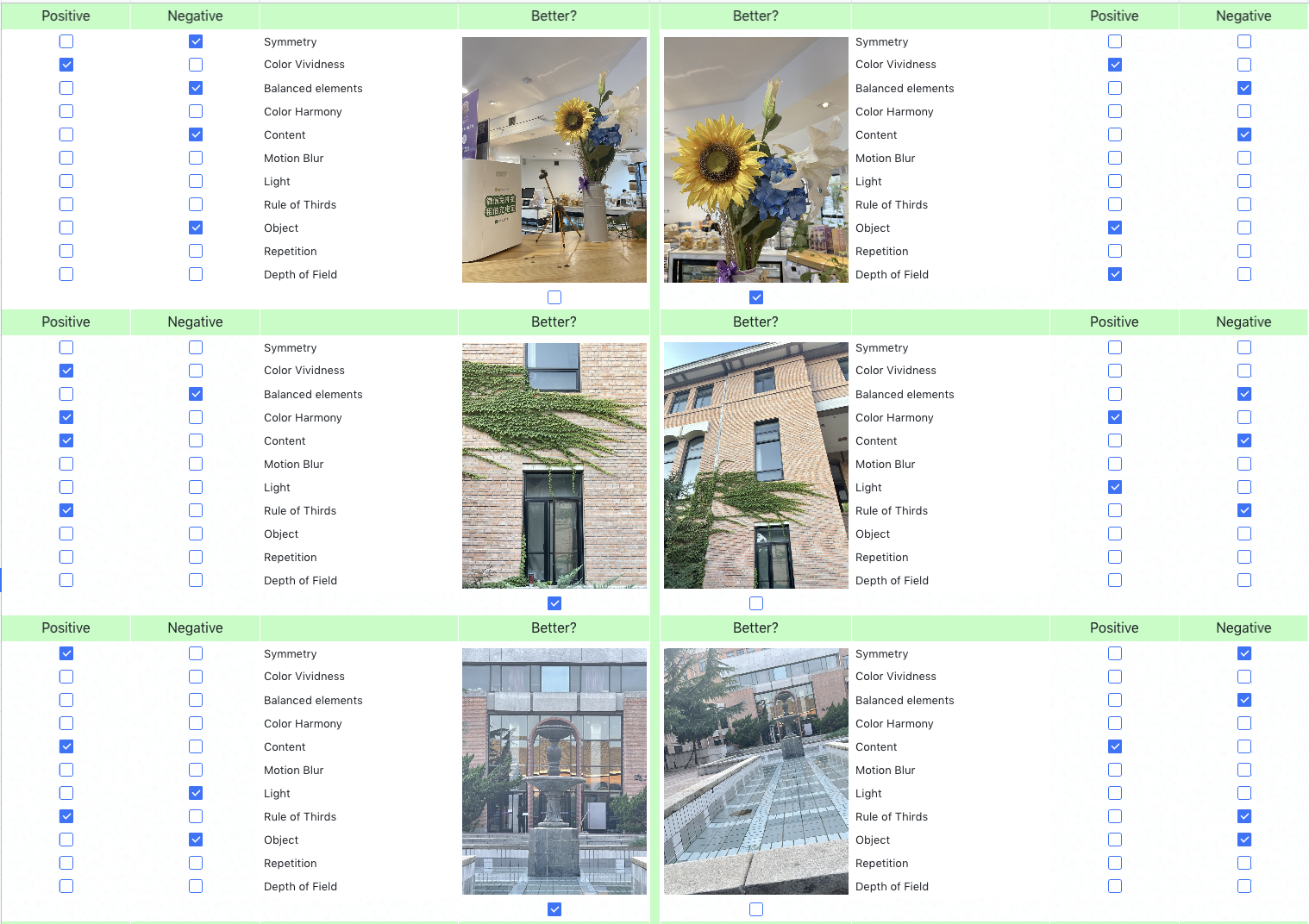}
    \includegraphics[width=\linewidth]{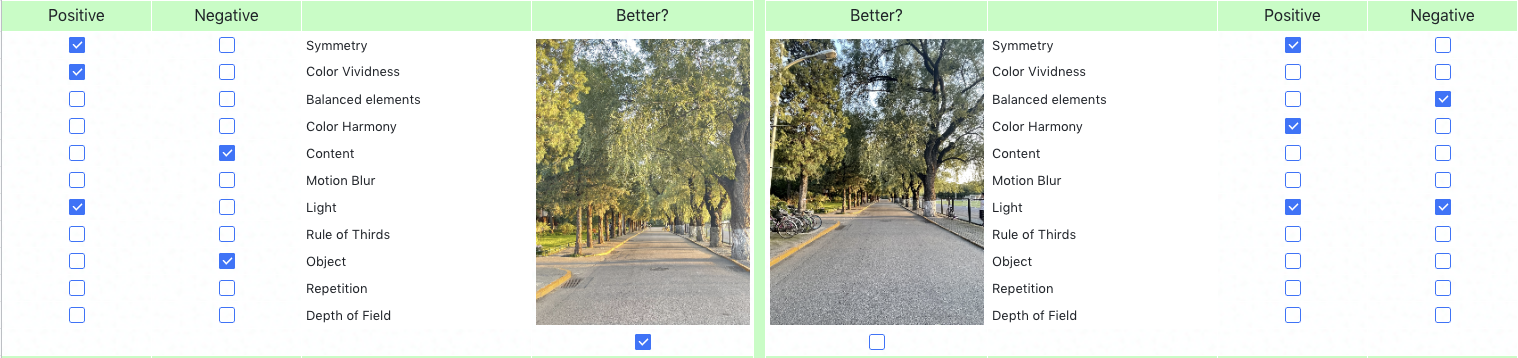}
    \caption{Part of a questionnaire finished by an expert.}
    \label{fig:form}
\end{figure}

\end{document}